\documentstyle[aps]{revtex}
\begin{document}
\title{{\rm DESY 97-239}\hspace{9.8cm}{\rm .}\\
{\rm December 1997}\hspace{9.6cm}\vspace{1.0in}\\
Linac-Ring Type Colliders: Fourth Way To TeV Scale}
\author{R. Brinkmann$^a$, A. K. \c {C}ift\c {c}i$^b$, S. Sultansoy$^{b,c}$, \c {S}.
T\"{u}rk\"{o}z$^b$, F. Willeke$^a$, \"{O}. Yava\c {s}$^b$,\qquad \qquad
\qquad \qquad \qquad \qquad \qquad \qquad M. Y\i {}lmaz$^d$}
\address{$^a$Deutches Elektronen Synchrotron, DESY\\
Notkestr. 85 \\
22607 Hamburg, GERMANY \\
$^{b}$Department of Physics, Faculty of Sciences\\
Ankara University \\
06100 Tando\u {g}an, Ankara, TURKEY\\
$^c$Institute of Physics, Academy of Sciences\\
H. Cavit Avenue\\
Baku, AZERBAIJAN\\
$^d$Department of Physics, Faculty of Arts and Sciences\\
Gazi University\\
06500 Be\c {s}evler, Ankara, TURKEY}
\maketitle

\begin{abstract}
The present status of suggested linac-ring type $ep$ and $\gamma p$
colliders is reviewed. The main parameters of these machines as well as $e$%
-nucleus and $\gamma $-nucleus colliders are considered. It is shown that
sufficiently high luminosities may be achieved with a reasonable improvement
of proton and electron beam parameters.\newpage\ 
\end{abstract}

\subsection{\bf Introduction}

The recent observation of $t$-quark at FNAL \cite{abe} is a triumph in the
realm of Standard Model (SM). In order to confirm the basic principles of
the SM, observation of the Higgs boson which gives the masses to the
fundamental particles is needed. Today, the SM explains practically all
experimental data on elementary particles and their interactions. However,
there are number of problems which have not been clarified by the SM. In
particular: i) Masses and mixings of fundamental fermions, ii) Existence of
fermion families and their number, iii) Arbitrary assignment of left-handed
particles to weak isospin doublets and right-handeds to singlets, iv) SM
does not really unify strong, weak and electromagnetic interactions, because
each of them are described by their own gauge groups (Appearance of photon
as a mixture of gauge bosons corresponding to weak hypercharge and third
component of weak isospin does not change the above statement). To solve
these problems, different approaches beyond the SM have been proposed:
extension of electroweak symmetry, GUT's, SUSY, preonic models, etc.
Generally, all extensions of the SM (with the possible exceptions of minimal
GUT's, namely, SU(5) and SO(10)) predict the rich spectrum of new particles
as well as new interactions at TeV scale.

It is known that the results from lepton-hadron collisions have been playing
a crucial role in particle physics. The establishment of the quark-parton
structure of the matter is the best example. If the new results \cite{Recent}
from HERA are verified we ought to review the research strategies in high
energy physics.

There are two well-known ways to reach TeV scale c.m. energies at
constituent level: the ring type proton-proton colliders and the linear $%
e^{+}e^{-}$ colliders (including $\gamma e$ and $\gamma \gamma $ options 
\cite{telnov}). Recently two more methods have been added to the list,
namely: the ring type $\mu ^{+}\mu ^{-}$ colliders \cite{palmer} and the
linac-ring type $ep$ colliders \cite
{grosse,sultansoy,tigner,zzaydin,vuop,brinkmann} (including $\gamma p$
option \cite{sultansoy,ciftci}).

In this paper we review the status of linac-ring type $ep$ and $\gamma p$
colliders. In the section B, main parameters of linac-ring type $ep$
colliders are discussed. The section C is devoted to main parameters of $%
\gamma p$ machines. Section D deals with $e$-nucleus and $\gamma $-nucleus
colliders. Finally, in section E, we briefly discuss physics goals of
colliders under consideration.

\subsection{{\bf Linac-ring type }$ep${\bf \ colliders }}

It is well-known that the energy achievable in an electron storage ring is
limited, because synchrotron radiation (SR) increases rapidly with the beam
energy ($\Delta E\sim E^4$). Even with the largest $e^{+}e^{-}$ storage ring
the beam energy can not be extended far beyond 100 GeV. In addition, the
luminosity in a storage ring is limited by the beam-beam interaction. The
way out of these problems is the linear collider scheme. Because, here the
SR losses are avoided and the beam-beam interaction limits are much weaker.
A number of high energy and high luminosity linear collider proposals are
under discussion (TESLA, NLC, JLC etc.) \cite{trines}. For the same reasons
new linac-ring type $ep$ colliders have been proposed in order to achieve
TeV scale c.m. energies at constituent level \cite
{grosse,sultansoy,tigner,zzaydin,vuop,brinkmann}. In this scheme the
electron beam from a linear accelerator (linac) is collided with a proton
beam stored in a ring. Essential advantage of the linac-ring type ep
colliders is closeness of electron and proton beam energies resulting with
good kinematics.

There are two reasons favoring a superconducting linear collider (TESLA) as
a source of $e$-beam for linac-ring type colliders. First of all the spacing
between bunches in conventional linac is only of the order of $ns.$ This
doesn't match with the bunch spacing in the proton ring. Also the pulse
length is much shorter than the ring circumference. Then in a conventional
linear collider, one can use only half of the machine, because the traveling
wave structures can accelerate only in one direction. There is no such
limitation for TESLA (standing wave cavities).

For these reasons we concentrate on HERA+TESLA and LHC+TESLA options. In
these cases following points and limitations should be considered:
Achievable luminosity is constrained by the electron beam power, limited
emittance of the proton beam and practical limits on focusing strength at
the interaction point. In addition, intra-beam scattering in proton beam
should be taken into account, since it causes emittance growth.

For round beams and equal transverse beam sizes for e and p at the crossing
point, the luminosity can be expressed as:

\begin{equation}
L_{ep}=\frac{n_en_pf_b\gamma _p}{4\pi \varepsilon _p^N\beta ^{*}}  \label{e}
\end{equation}
where $\varepsilon _p^N$ is normalized proton beam emittance, $n_e$ and $n_p$
are the numbers of electrons and protons per bunch, $f_b$ is collision
frequency, $\gamma _p$ is proton Lorentz factor and $\beta ^{\star }$ is
amplitude function at interaction point.

Once the electron beam energy is chosen, the total electron beam current $%
I_e=n_e\cdot e\cdot f_b$ is limited by the allowed electron beam power ($%
I_e=P_e/E_e$). As follows from Eqn.(1), $L_{ep}$ is independent of $n_e$ and 
$f_b$ as long as their product is constant. With the consideration of power
limitation, luminosity is given by \cite{tigner}

\begin{equation}
L=1.66\times 10^{31}s^{-1}cm^{-2}\frac{n_p}{10^{11}}\cdot \frac{P_e}{10^8Watt%
}\cdot \frac{0.3TeV}{E_e}\cdot \frac{\gamma _p}{1000}\cdot \frac{10^{-6}}{%
\varepsilon _p^N}\cdot \frac{10cm}{\beta ^{*}}  \label{f}
\end{equation}
Center of mass energy and luminosity values for HERA+TESLA and LHC+TESLA
options are shown in Table I. Here, $n_p=3\times 10^{11}$, $\varepsilon
_p^N=0.8\times 10^{-6}m$, $\beta ^{*}=20\,cm,$ $P_e=40$ $MW$ for 0.3 TeV and 
$P_e=70$ $MW$ for 0.96 TeV are used in calculations. Appropriate parameters
of corresponding TESLA beams are given in Table II.

In a recent paper \cite{brinkm} HERA+TESLA\ option is studied in more
detail. At this study luminosity limitation, interaction region layout,
intra-beam scattering (IBS) growth times and proton beam cooling for this
collider are investigated. It is shown that, with some modifications of
accelerating gradient and RF-system power, beam pulse current can be
increased by a factor of 2.5. Luminosity for this option ($%
E_p=820GeV,E_e=250GeV$) has been calculated as $1.3\cdot
10^{31}cm^{-2}s^{-1}.$ In addition, for $\varepsilon _x\varepsilon _y=const$%
, using flat proton beam ($\varepsilon _y=0.1\varepsilon _x$) IBS growth
time can be taken as 2 hours while for $\varepsilon _y=\varepsilon _x$ it is
less than one hour. For higher luminosity, emittance cooling of proton beams
at very high energies must be managed by advanced cooling methods as
optical, stochastic or electron cooling. To satisfy an emittance doubling
time to be not smaller than 5h, SPS experience shows that tune shift
parameter is better to be limited to $\Delta Q=4\cdot 10^{-3}$ \cite{SPS}.
Lower limit on $\beta ^{*}$ is given by the p-beam bunch length. This
limitation can be overcome by applying dynamic focusing scheme, where the
p-beam waist travels with the e-bunch during collision \cite{brinkmann}.

\subsection{Linac-ring type $\gamma $p colliders}

Earlier, the idea of obtaining high energy photon beams by Compton
backscattering of laser light off a beam of high energy electrons was
considered for $\gamma e$ and $\gamma \gamma $ colliders\cite{ginzburg}. The
same method can be used for constructing $\gamma p$ colliders on the base of
linac-ring type $ep$ machines. A $\gamma p$ collider option will open
important horizons to the TeV scale physics. At the standard ep colliders
proton can be probed by Weizs\"{a}cker-Williams photons. When we compare
Weizs\"{a}cker-Williams photon spectrum with real photon spectrum obtained
from backscattering, we easily see advantage of the real photon beam since
real photons are generated mostly on high energy end of the energy scale
while Weizs\"{a}cker-Williams photons are generated mostly on low energy
end. The second advantage of linac-ring type $\gamma p$ collider is good
kinematics, that is, the energy of the $\gamma $ beam is close to the energy
of the valence quark. Therefore, events after collisions can be observed at
wider angles. The third advantage of the linac-ring type $\gamma p$ collider
is high polarization of $\gamma $ beam.

For the main parameters of the linac-ring type $\gamma p$ colliders, we will
use electron and proton beam parameters that are given in the preceding
section.

Taking into account the effects of the distance between the conversion
region and the collision point, the luminosity is given as 
\begin{equation}
L_{\gamma p}=\int_0^{\omega _{max}}\frac{dL_{\gamma p}}{d\omega }d\omega
\label{iii}
\end{equation}
where differential luminosity is given by

\begin{equation}
\frac{dL_{\gamma p}}{d\omega }=\frac{f(\omega )n_\gamma n_pf_\gamma }{2\pi
(\sigma _e^2+\sigma _p^2)}exp[-z^2\Theta _\gamma ^2(\omega )/2(\sigma
_e^2+\sigma _p^2)]  \label{iv}
\end{equation}
where $n_\gamma =n_e$ (conversion is considered as one to one), $f_\gamma
=f_b$, $\sigma _e=\sqrt{\varepsilon _e^N\beta _e^{*}/\gamma _e}$, $\sigma _p=%
\sqrt{\varepsilon _p^N\beta ^{*}/\gamma _p}$. Here, $\omega $ is high energy
photon's energy, $z$ is the distance between conversion region and collision
point, $\Theta _\gamma (\omega )$ is the angle between high energy photons
and the electron beam direction. This angle is given by (for small $\Theta
_\gamma $)

\begin{equation}
\Theta _\gamma (\omega )=\frac{m_e}{E_e}\sqrt{\frac{E_e\kappa }\omega
-(\kappa +1)}  \label{v}
\end{equation}
where $\kappa =4E_e\omega _0/m_e^2$, $\omega _0$ is laser photon energy. In
order to avoid $e^{+}e^{-}$ pair creation in the conversion region, $\kappa $
should be less than $4.83$. In Eqn.(\ref{iv}), $f(\omega )$ is the
normalized differential Compton cross section \cite{ginzburg}:

\begin{equation}
f(\omega )=\frac 1{E_b\sigma _c}\frac{2\pi \alpha ^2}{\kappa m_e^2}[\frac 1{%
1-y}+1-y-4r(1-r)+\lambda _e\lambda _0r\kappa (1-2r)(2-y)]  \label{vi}
\end{equation}
where $y=\omega /E_e$, $r=y/[\kappa (1-y)]$. The total Compton cross section
is

\begin{eqnarray}
\sigma _c &=&\sigma _c^0+\lambda _e\lambda _0\sigma _c^1  \nonumber \\
\sigma _c^0 &=&{\frac{\pi \alpha ^2}{\kappa m_e^2}}[(2-\frac 8\kappa -\frac{%
16}{\kappa ^2})\ln (\kappa +1)+1+\frac{16}\kappa -\frac 1{(\kappa +1)^2}]
\label{vii} \\
\sigma _c^1 &=&\frac{\pi \alpha ^2}{\kappa m_e^2}[(2+\frac 4\kappa )\ln
(\kappa +1)-5+\frac 2{\kappa +1}-\frac 1{(\kappa +1)^2}]  \nonumber
\end{eqnarray}
In the equations above, $\lambda _e$ and $\lambda _0$ are helicities of
electron and laser photon. As one can see from Eqn.(\ref{v}), high energy
photons with maximal energy 
\begin{equation}
\omega _{max}=E_e\frac \kappa {\kappa +1}=0.83E_e  \label{viii}
\end{equation}
move in the direction of electron beam.

In this paper we deal with equal electron and proton beam sizes, therefore
Eqn.(\ref{iv}) takes form 
\begin{equation}
\frac{dL_{\gamma p}}{d\omega }=f(\omega )L_{ep}\exp [-z^2\Theta _\gamma
^2(\omega )\gamma _p/4\varepsilon _p^N\beta ^{*}]  \label{ix}
\end{equation}

In Figures 1 and 2, the dependence of luminosity on distance between the
conversion region and the collision point is plotted for three different
choices of laser photons and electron beam helicities for HERA+TESLA and
LHC+TESLA, respectively. Minimum wavelengths of the laser (corresponding to $%
\kappa =4.83$) for HERA+TESLA and LHC+TESLA are estimated as $1.2\,\mu m$
and $3.8\,\mu m$, respectively. Mirror systems for the multiple use of laser
pulses for the conversion of a number of successive electron bunches are
necessary in the case of multibunch structure of linac pulses. This point
raises from the fact that lasers with sufficiently high power per pulse seem
to be not available in the near future for pulse rates much higher than 100
Hz.

Figure 3 shows luminosity distributions of HERA+TESLA depending on $\gamma p$
invariant mass $W_{\gamma p}=2\sqrt{E_\gamma E_p}$ at $z=6\,$m. Again the
advantage of opposite helicities of laser photons and electron beam is
obvious. Similar plots hold for LHC+TESLA. Interpretation of these figures
leads to some important conclusions: Luminosity slowly decreases with the
distance between the conversion region and the collision point. Opposite
helicity values for laser and electron beams are advantageous for photon
spectrum. Additionally, a better monochromatization for high energy gamma
beam can be achieved by increasing the distance $z$. Also mean helicity of
the high energy gamma beam approaches to unity with increasing distance.

\subsection{e-Nucleus and $\gamma $-Nucleus Colliders}

The linac-ring type $ep$ and $\gamma p$ colliders could be used with some
modification as e-nucleus and $\gamma $-nucleus machines \cite{ichep96} to
investigate the properties of nuclei at TeV scale region. Indeed, LHC will
give opportunity to accelerate nuclei too, and the acceleration of nuclei in
HERA ring is under discussion. Therefore, we estimate basic parameters for
such machines with the examples of carbon and lead beams (Table III). When
it comes to electron beam, it is considered to be produced by TESLA (Table
II).

Center of mass energy and luminosity values of e-nucleus and $\gamma $%
-nucleus collisions are presented in Table IV. Here we take into account
that the energy of converted photons is restricted by the condition given in
Eqn. (8). Then, the luminosity of e-nucleus collisions is calculated using
Eqns. (1) and (2) with appropriate substitution of proton beam parameters
with nucleus beam parameters. It is clear that $L_{\gamma \text{-nucleus}%
}\approx L_{\text{e-nucleus}}$ at $z=0m$. The dependence of luminosity on
distance between the conversion region and collision point is similar to $%
\gamma p$ option: at $z=10m$ luminosity value reduces by factor $\sim 1.5$
comparing to that at $z=0m$.

We also present the values of $\gamma $-nucleon luminosities and
luminosities per collision. Multiplying last ones by extrapolated cross
sections ($2\times 10^{-27}$ cm$^2$ for $\gamma $-C and $4\times 10^{-26}$ cm%
$^2$ for $\gamma $-Pb collisions) one obtain the number of events per
collision and this number is less than one.

Finally, by applying dynamic focusing scheme \cite{brinkmann} one can use
essentially lower values of $\beta ^{*}$ and therefore achieve several times
higher luminosities. Further increase of the luminosity might be achieved if
effective method of cooling of nuclei beams is invented.

\subsection{Physics at TeV Energy $ep$ and $\gamma p$ Colliders}

Although physics search programs of new $ep$ and $\gamma p$ colliders are
much less developed than those of LHC, NLC and $\mu ^{+}\mu ^{-}$ collider,
a lot of papers on this subject were published during the last decade. The
physics at UNK$+$VLEPP based $ep$ and $\gamma p$ colliders were considered
in \cite{ihep87} and \cite{ihep88,ijmp}, correspondingly. Resonant
production of excited quarks at $\gamma p$ colliders was investigated in 
\cite{jikia}. Reference \cite{boos} dealt with physics at future $\gamma
\gamma $, $\gamma e$ and $\gamma p$ colliders. Wino production at HERA+LC
based $\gamma p$ collider was considered in \cite{bf}. Today, the main
activity on this subject is concentrated in Ankara University HEP group
[23-34]. In Reference \cite{aydin} physics search potential of HERA+LC based 
$\gamma p$ collider had been reviewed. Recently, Higgs boson production at
LHC+LEP based $\gamma p$ collider has been studied in \cite{cheung} and \cite
{lm}, however their results should be recalculated because (as argued in 
\cite{ciftci}) $\gamma p$ colliders can be constructed only on the base of
linac-ring type $ep$ machines.

\subsubsection{Physics at Linac-Ring Type $ep$ Colliders}

This topic was sufficiently developed during preparation of HERA and study
of LHC+LEP physics search potential. The linac-ring type machines will give
opportunity to investigate appropriate phenomena at \smallskip\ 

$\circ $ higher center of mass energies,

$\circ $ better kinematic conditions. \smallskip\ 

The situation is illustrated by the following table:

\[
\begin{tabular}{lllll}
& \quad HERA & \quad LHC+LEP & \quad HERA+TESLA & \quad LHC+TESLA \\ 
$\sqrt{s_{ep}}$, TeV & \quad 0.3 & \quad 1.2 & \quad 1(2.4) & \quad 2.6(6.5)
\\ 
$E_e/E_p$ & \quad 1/30 & \quad 1/120 & \quad 1/4 & \quad 1/5
\end{tabular}
\]

Let us remind that confirmation of recent results \cite{Recent} from HERA
will favor new $ep$ machines. Physics search program of HERA+TESLA based $ep$
collider is considered in \cite{roeck}.

\subsubsection{Physics at $\gamma p$ Colliders}

Below we illustrate physics search potential of future $\gamma p$ machines.
As samples we use HERA+TESLA(1 TeV$\times $0.3 TeV) with $L_{\gamma
p}^{int}=500pb^{-1}$ and LHC+TESLA(7 TeV$\times $1.5 TeV) with $L_{\gamma
p}^{int}=5fb^{-1}$.

\paragraph{SM Physics\protect\\$\quad $}

$\bullet $ Total cross-section at TeV scale can be extrapolated from
existing low energy data as $\sigma (\gamma p\rightarrow hadrons)\sim
100\div 200\mu b$, which corresponds to $\sim 10^{11}$ hadronic events per
working year \smallskip\ 

$\bullet $ Two-jet events (large $p_t$)

\qquad HERA+TESLA: 10$^4$ events with $p_t$%
%TCIMACRO{\TEXTsymbol{>}}
%BeginExpansion
\mbox{$>$}
%EndExpansion
100 GeV

\qquad LHC+TESLA: 10$^4$ events with $p_t$%
%TCIMACRO{\TEXTsymbol{>}}
%BeginExpansion
\mbox{$>$}
%EndExpansion
500 GeV \smallskip\ 

$\bullet $ $t\overline{t}$ pair production

\qquad HERA+TESLA: 10$^3$ events per year

\qquad LHC+TESLA: 10$^5$ events per year \smallskip\ 

$\bullet $ $b\overline{b}(c\overline{c})$ pair production

\qquad HERA+TESLA: 10$^8$ events

\qquad LHC+TESLA: 10$^9$ events

the region of extremely small $x_g\sim 10^{-6}\div 10^{-7}$ can be
investigated \smallskip\ 

$\bullet $ $W$ production

\qquad HERA+TESLA: 10$^5$ events

\qquad LHC+TESLA: 10$^6$ events

$\Delta \kappa _W$ can be measured with accuracy of 0.01 \smallskip\ 

$\bullet $ Higgs boson production ($\gamma p\rightarrow WH+X)$

\qquad HERA+TESLA: 20 events at $m_H=$ 100 GeV

\qquad LHC+TESLA: 1000 events at $m_H=$ 100 GeV and 100 events at $m_H=$ 300
GeV \smallskip\ 

$\bullet $ Fourth SM\ family quarks (discovery limits for 100 events per
year)

\qquad HERA+TESLA: $m_{u_4}=250$ GeV, $m_{d_4}=200$ GeV

\qquad LHC+TESLA: $m_{u_4}=1000$ GeV, $m_{d_4}=800$ GeV

\paragraph{Beyond the SM Physics\protect\\}

Below we present discovery limits for 100 events per year: \smallskip\ 

$\bullet $ $\gamma p$ colliders are ideal machines for $u^{*},$ $d^{*}$ and $%
Z_8$ search

\qquad HERA+TESLA: $m_{u^{*}}=0.9$ TeV, $m_{d_{}^{*}}=0.7$ TeV, $m_{Z_8}=0.7$
TeV

\qquad LHC+TESLA: $m_{u^{*}}=5$ TeV, $m_{d_{}^{*}}=4$ TeV, $m_{Z_8}=4$ TeV %
\smallskip\ 

$\bullet $ single $l_q$ production

$\qquad $HERA+TESLA: 0.7 TeV

$\qquad $LHC+TESLA: 3 TeV$\quad $ \smallskip\ 

$\bullet $ pair $l_q$ production

$\qquad $HERA+TESLA: 0.3 TeV

$\qquad $LHC+TESLA: 1.7 TeV \smallskip\ 

$\bullet $ pair squark production

$\qquad $HERA+TESLA: 0.25 TeV

$\qquad $LHC+TESLA: 0.8 TeV \smallskip\ 

$\bullet $ associative gaugino-squark production ($m_{\tilde{G}}+m_q$)

$\qquad $HERA+TESLA: 0.5 TeV

$\qquad $LHC+TESLA: 2.0 TeV \smallskip\ 

It is clear that physics search potential of linac ring type ep and $\gamma $%
p colliders is comparable with and complementary to that of basic hadron and
lepton colliders. Upgraded Tevatron, NLC with $\sqrt{s}=0.5TeV$ and
HERA+TESLA will give possibility to investigate in details 0.5TeV scale.
Similarly LHC, NLC with $\sqrt{s}=2TeV$ , $\mu ^{+}\mu ^{-}$ with $\sqrt{s}%
=4TeV$ and LHC+TESLA will give opportunity to investigate TeV scale physics
in the best manner by use of all possible colliding beams.

\subsubsection{Physics at $\gamma $-nucleus colliders}

Center of mass energy of LHC+TESLA based $\gamma $-nucleus collider
corresponds to $E_\gamma \sim $ PeV in the lab. system. At this energy range
cosmic ray experiments have a few events per year, whereas $\gamma $-nucleus
collider will give few billions events. Very preliminary list of physics
goals contains: \smallskip\ 

$\circ $ total cross-sections to clarify real mechanism of very high energy $%
\gamma $-nucleus interactions

$\circ $ investigation of hadronic structure of the photon in nuclear medium

$\circ $ according to the VMD, proposed machine will be also $\rho $-nucleus
collider

$\circ $ formation of the quark-gluon plasma at very high temperatures but
relatively low nuclear densities

$\circ $ gluon distribution at extremely small $x_g$ in nuclear medium ($%
\gamma A\rightarrow Q\overline{Q}+X$)

$\circ $ investigation of both heavy quark and nuclear medium properties ($%
\gamma A\rightarrow J/\Psi (\Upsilon )+X$, $J/\Psi (\Upsilon )\rightarrow
l^{+}l^{-}$)

$\circ $ existence of multi-quark clusters in nuclear medium and few-nucleon
correlations.

\subsection{Conclusion}

There are strong arguments favoring that the rich spectrum of new particles
and/or interactions will manifest themselves at TeV scale. An exploration of
this scale at constituent level will require all possible types of colliding
beams. Today, work on physics search programs and machine parameters for
future hadron and lepton colliders is quite advanced, whereas those for
linac-ring type lepton-hadron colliders need an additional R\&D. In machine
aspects: it seems possible to increase luminosity up to $%
10^{31}cm^{-2}s^{-1} $ for HERA+TESLA and up to $10^{32}cm^{-2}s^{-1}$ for
LHC+TESLA by the development in this field (dynamic focusing, cooling
technics, etc.). For physics goals: more detailed investigations of all
possible processes are needed.

\begin{center}
{\bf Acknowledgements}
\end{center}

Authors are grateful to A. \c {C}elikel, D. Trines and B.H. Wiik for useful
discussions.

This work is supported in part by Turkish State Planning Organization (DPT)
under the grant number 97K-120-420.

\newpage

%TCIMACRO{\TeXButton{Table I.}{\begin{table}[tbp] \centering}}
%BeginExpansion
\begin{table}[tbp] \centering
%EndExpansion
\caption{ Center of mass energy and Luminosity values for HERA+TESLA 
and  LHC+TESLA Colliders.  
\label{Table 1.}} 
\begin{tabular}{|l|l|l|l|l|}
Collider & $E_p(TeV)$ & $E_e(TeV)$ & $\sqrt{s}(TeV)$ & $%
L(10^{31}cm^{-2}s^{-1})$ \\ \hline
HERA+TESLA($ep$) & 0.82 & 0.30 & 1.00 & 1.20 \\ 
LHC+TESLA($ep$) & 7 & 0.96 & 5.18 & 6.00
\end{tabular}
%TCIMACRO{\TeXButton{E}{\end{table}}}
%BeginExpansion
\end{table}
%EndExpansion

%TCIMACRO{\TeXButton{Table II.}{\begin{table}[tbp] \centering}}
%BeginExpansion
\begin{table}[tbp] \centering
%EndExpansion
\caption{Some parameters of Superconducting  $e^{-}$ Linac TESLA for
0.3 and 0.96 TeV versions.
\label{Table 2.}} 
\begin{tabular}{|c|c|c|}
Electron Energy, (GeV) & 300 & 960 \\ \hline
Number of Electrons per Bunch, (10$^{10})$ & 1.8 & 0.9 \\ \hline
Beam Power, (MW) & 40 & 70 \\ \hline
Bunch Length, mm & 1 & 1 \\ \hline
Invariant Emittance, (10$^{-6}m)$ & 2200 & 825 \\ \hline
Beta Function at IP, mm & 200 & 200 \\ \hline
Bunch Spacing, ns & 192 & 100 \\ \hline
RF Frequency, MHz & 1300 & 1300 \\ \hline
Acceleration Gradient,MV/m & 15 & 25 \\ \hline
Pulse Length, $\mu s$ & 1000 & 1000 \\ \hline
Repetition Rate, Hz & 10 & 6
\end{tabular}
%TCIMACRO{\TeXButton{E}{\end{table}}}
%BeginExpansion
\end{table}
%EndExpansion

%TCIMACRO{\TeXButton{Table III.}{\begin{table}[tbp] \centering}}
%BeginExpansion
\begin{table}[tbp] \centering
%EndExpansion
\caption{ Main parameters of nucleus beams } \label{Table 3.} 
\begin{tabular}{|c|c|c|c|}
PARAMETERS & C(HERA) & Pb(HERA) & Pb(LHC) \\ \hline
Maximum beam energy (TeV) & 4.9 & 67 & 574 \\ \hline
Particles per bunch, $n$ $(10^8)$ & 80 & 0.5 & 0.94 \\ \hline
Normalized emittance, $\varepsilon ^N$ (mm-mrad) & 1.25 & 2 & 1.4 \\ \hline
Bunch length, $\sigma _z$ (cm) & 15 & 15 & 7.5 \\ \hline
Bunch spacing (ns) & 192 & 192 & 100
\end{tabular}
%TCIMACRO{\TeXButton{E}{\end{table}}}
%BeginExpansion
\end{table}
%EndExpansion

%TCIMACRO{\TeXButton{Table V.}{\begin{table}[tbp] \centering}}
%BeginExpansion
\begin{table}[tbp] \centering
%EndExpansion
\caption{Center of mass energies and luminosities of e-nucleus and 
$\gamma \text{-}nucleus$ collisions (at $\beta=0.2$m).} \label{Table 5.} 
\begin{tabular}{|c|c|c|c|}
& C(HERA) & Pb(HERA) & Pb(LHC) \\ \hline
$\sqrt{s}$ (e-nucleus) (TeV) & 2.4 & 9.0 & 47 \\ \hline
$\sqrt{s}^{\max }$ ($\gamma $-nucleus) (TeV) & 2.2 & 8.1 & 43 \\ \hline
$L$ (in $10^{28}$ cm$^{-2}$s$^{-1}$) & 10 & 0.02 & 0.4 \\ \hline
$L\cdot A$ (in $10^{30}$ cm$^{-2}$s$^{-1}$) & 1.2 & 0.04 & 0.8 \\ \hline
$L$/coll (in $10^{23}cm^{-2}$) & 20 & 0.04 & 0.7
\end{tabular}
%TCIMACRO{\TeXButton{E}{\end{table}}}
%BeginExpansion
\end{table}
%EndExpansion

\break 

\begin{center}
%TCIMACRO{\TeXButton{Figure I}{\begin{figure}[tbp] \centering}}
%BeginExpansion
\begin{figure}[tbp] \centering
%EndExpansion
\input{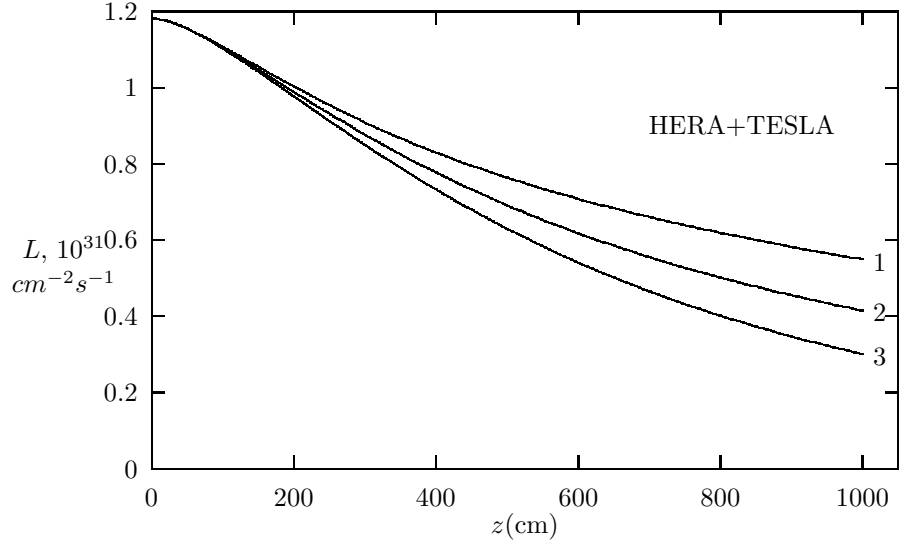}\medskip\ 
\caption{Dependence of the luminosity (in units of $10^{31}cm^{-2}s^{-1}$) on 
distance between the conversion region and the collision point for HERA+TESLA. 
Curves 1, 2 and 3 correspond to $\lambda _e\lambda _0=-1$, 
$\lambda _e\lambda _0=0$ and $\lambda _e\lambda _0=1$, respectively.
\label{Figure 1}}%
%TCIMACRO{\TeXButton{E}{\end{figure}}}
%BeginExpansion
\end{figure}
%EndExpansion
\smallskip

\bigskip\ 

%TCIMACRO{\TeXButton{Figure II}{\begin{figure}[tbp] \centering}}
%BeginExpansion
\begin{figure}[tbp] \centering
%EndExpansion
\input{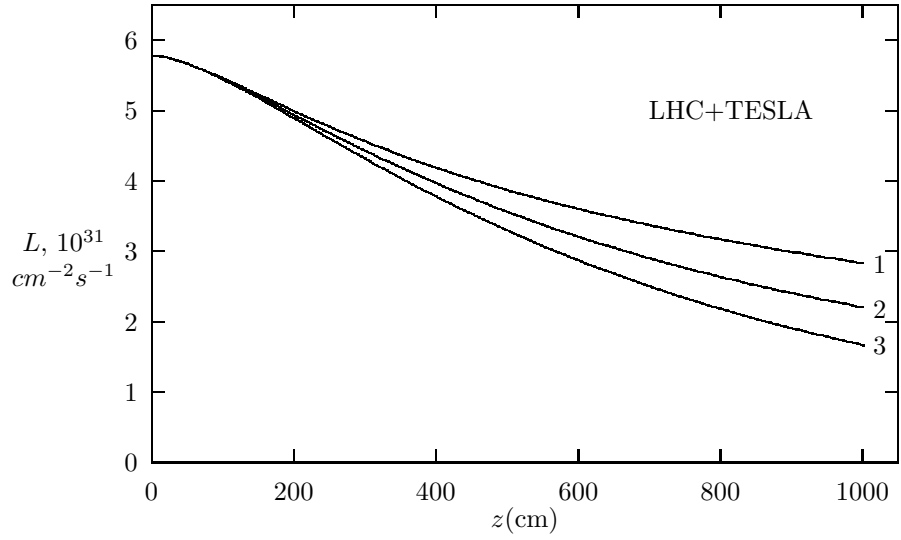}\medskip\ 
\caption{Dependence of the luminosity (in units of $10^{31}cm^{-2}s^{-1}$) on 
distance between the conversion region and the collision point for LHC+TESLA. 
Curve numeration is same as in the Fig. 1.
\label{Figure 2}}%
%TCIMACRO{\TeXButton{E}{\end{figure}}}
%BeginExpansion
\end{figure}
%EndExpansion
\smallskip

\bigskip\ 

%TCIMACRO{\TeXButton{Figure III}{\begin{figure}[tbp] \centering}}
%BeginExpansion
\begin{figure}[tbp] \centering
%EndExpansion
\input{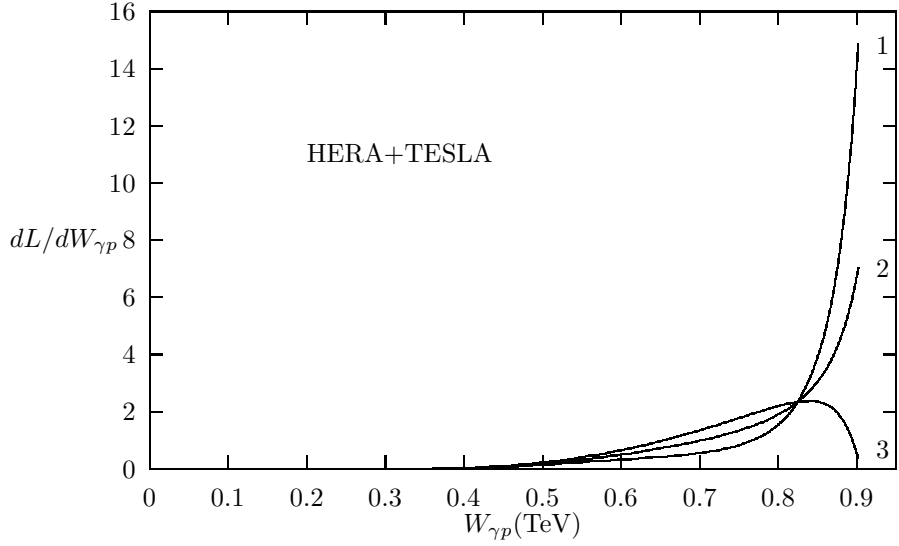}\medskip\ 
\caption{Luminosity distributions (in units of $10^{31}cm^{-2}s^{-1}TeV^{-1}$) depending 
on $\gamma p$ invariant mass at $z=6$ m for HERA+TESLA. Curve numeration is same as 
in the Fig. 1.
\label{Figure 3}}%
%TCIMACRO{\TeXButton{E}{\end{figure}}}
%BeginExpansion
\end{figure}
%EndExpansion
\smallskip
\end{center}

\end{document}